\documentclass{ws-procs961x669}            % default, citations in superscript
\newcommand{\EQ}{\begin{equation}}
\newcommand{\EN}{\end{equation}}
\newcommand{\EQA}{\begin{eqnarray}}
\newcommand{\ENA}{\end{eqnarray}}

\newcommand{\Eq}[1]{Eq.~(\ref{#1})}

\newcommand{\Sec}[1]{Sect.~\ref{#1}}

\newcommand{\Secss}[2]{Sects.~\ref{#1}--\ref{#2}}

\newcommand{\Fig}[1]{Figure~\ref{#1}}
\newcommand{\FFig}[1]{Figure~\ref{#1}}

\newcommand{\Tab}[1]{Table~\ref{#1}}

\newcommand{\bra}[1]{\langle #1\rangle}

{}
{}
{}

{}
{}
\newcommand{\meanEMF}{\overline{\mbox{\boldmath ${\cal E}$}}{}}{}
{}
{}
{}
{}
{}
{}
{}
{}
{}
{}
{}
{}
{}
{}
\newcommand{\meanUU}{\overline{\bm{U}}}

{}

{}

{}
{}
{}

\newcommand{\tildeh}{\tilde{h}}

\newcommand{\tildeee}{\tilde{\mathbf{e}}}
\newcommand{\tildeAA}{\tilde{\mathbf{A}}}

\newcommand{\tildeA}{\tilde{A}}

\newcommand{\eee}{\hat{\mbox{\boldmath $e$}} {}}
\newcommand{\nnn}{\hat{\bm{n}}}

\newcommand{\meanAA}{{\overline{\bm{A}}}}
\newcommand{\meanBB}{{\overline{\bm{B}}}}
\newcommand{\meanJJ}{{\overline{\bm{J}}}}

\newcommand{\vv}{\mbox{\boldmath $v$} {}}

\newcommand{\kk}{\bm{k}}

\newcommand{\pp}{\bm{p}}

\newcommand{\aaaa}{\bm{a}}
\newcommand{\jj}{\bm{j}}
\newcommand{\rr}{\bm{r}}
\newcommand{\grav}{\bm{g}}
\newcommand{\bb}{\bm{b}}
\newcommand{\BB}{\bm{B}}

\newcommand{\EE}{\bm{E}}
\newcommand{\JJ}{\bm{J}}
\newcommand{\oo}{\bm{\omega}}

\newcommand{\eell}{\bm{\ell}}
\newcommand{\AAA}{\bm{A}}

\newcommand{\UU}{\bm{U}}

\newcommand{\uu}{\bm{u}}

\newcommand{\sss}{\mbox{\boldmath $s$} {}}
\newcommand{\SSS}{\bm{S}}

\newcommand{\FFF}{\mbox{\boldmath ${\cal F}$} {}}

\newcommand{\nab}{{\bm{\nabla}}}

\newcommand{\OO}{\bm{\Omega}}

\newcommand{\ii}{{\rm i}}

\newcommand{\dd}{{\rm d} {}}

\newcommand{\const}{{\rm const}  {}}

\def\degr{\hbox{$^\circ$}}
\def\la{\mathrel{\mathchoice {\vcenter{\offinterlineskip\halign{\hfil
$\displaystyle##$\hfil\cr<\cr\sim\cr}}}
{\vcenter{\offinterlineskip\halign{\hfil$\textstyle##$\hfil\cr<\cr\sim\cr}}}
{\vcenter{\offinterlineskip\halign{\hfil$\scriptstyle##$\hfil\cr<\cr\sim\cr}}}
{\vcenter{\offinterlineskip\halign{\hfil$\scriptscriptstyle##$\hfil\cr<\cr\sim\cr}}}}}

\def\Imag{\mbox{\rm Im}}

\def\EM{E_{\rm M}}

\def\cs{c_{\rm s}}

\def\xiM{\xi_{\rm M}}

\def\kf{k_{\rm f}}

\def\HM{H_{\rm M}}
\def\EM{E_{\rm M}}

\def\kB{k_{\rm B}}

\def\kB{k_{\rm B}}

\def\Brms{B_{\rm rms}}

\def\etat{\eta_{\rm t}}

\newcommand{\nT}{\,{\rm nT}}
\newcommand{\T}{\,{\rm T}}
\newcommand{\G}{\,{\rm G}}

\newcommand{\uG}{\,\mu{\rm G}}

\newcommand{\cm}{\,{\rm cm}}

\newcommand{\m}{\,{\rm m}}
\newcommand{\km}{\,{\rm km}}

\newcommand{\Mm}{\,{\rm Mm}}

\newcommand{\pc}{\,{\rm pc}}

\newcommand{\Mpc}{\,{\rm Mpc}}

\newcommand{\AU}{\,{\rm AU}}

\begin{document}
\title{Chirality in Astrophysics}

\author{Axel Brandenburg}

\address{Nordita, KTH Royal Institute of Technology and Stockholm University, and\\
The Oskar Klein Centre, Department of Astronomy, Stockholm University, Stockholm, Sweden\\
$^*$E-mail: brandenb@nordita.org, \today, $ $Revision: 1.65 $ $\\
\url{https://www.nordita.org/~brandenb}}

\begin{abstract}
Chirality, or handedness, enters astrophysics in three distinct ways.
First, magnetic field and vortex lines tend to be helical and have a systematic
twist in the northern and southern hemispheres of a star or a galaxy.
Helicity is here driven by external factors.
Second, chirality can also enter at the microphysical level and can then be traced
back to the parity-breaking weak force.
Third, chirality can arise spontaneously, but this requires not
only the presence of an instability, but also the action of nonlinearity.
Examples can be found both in magnetohydrodynamics and in astrobiology,
where homochirality among biomolecules probably got established at the
origin of life.
In this review, all three types of chirality production will be explored
and compared.
\end{abstract}

\keywords{Magnetic helicity; chiral magnetic effect; gravitational waves; homochirality}

\bodymatter

\vspace{1cm}

\section{Introduction}

Chirality, or handedness, plays important roles in many different fields
of astrophysics, including astrobiology.
There are three distinct types of chirality production:
(i) driven chirality due to external factors,
(ii) driven chirality due to intrinsic properties, and
(iii) spontaneous chirality production due to instability and nonlinearity.
The primary applications for these three types of chirality production
are rather distinct, but one can find unifying circumstances under which
the different types can be demonstrated and compared.
One such circumstance is given by the presence of magnetic fields.

Magnetic fields can experience twisting that makes them helical, but the
central question is what determines the sign of this twist---especially
if it is a systematic one, always being in the same sense.
When there are extrinsic or intrinsic factors such as the combination
of rotation and stratification in a star, or intrinsic factors such as
the presence of fermions of one of two handednesses, the answer is in
principal clear.
However, there can also be spontaneous helicity production, where the
sign depends ultimately on chance, so both signs are possible under
almost identical conditions.
An example that we discuss at the end of this review is in the field
of magnetohydrodynamics (MHD), where a magnetic field in a stratified
atmosphere that exhibits a magnetic buoyancy instability where, in the
end, once nonlinearity plays a role, one particular handedness dominates
to nearly hundred percent.
It is this example that, in a broad sense, also carries over to
astrobiology and the origin of life, where one particular chirality
of biomolecules eventually dominates and leads, to what is known as
homochirality.
Magnetic fields are probably not involved in the origin of life,
but there are simple mathematical analogies in both cases.

In \Tab{Toverview}, we summarize the three types of chirality production
under the ``what'' column or category.
The ``where'' category lists some specific examples, and the ``how''
category highlights some specific techniques for measuring chirality
for those three types.

\begin{table}[t]\tbl{
Summary of three types of chirality production in astrophysics.
}{\begin{tabular}{@{}lll@{}}
\toprule
What & Where & How \\
\colrule
Helicity, driven by        & stars, planets, & magnetic field from Zeeman effect, \\
stratification \& rotation & galaxies        & polarization, in situ in solar wind \\
\hline
Fermion asymmetry,         & entire universe & polarization patters, photon arrival \\
axions                     &                 & direction statistics, circularly \\
                           &                 & polarized gravitational waves \\
\hline
Spontaneous chirality      & astrobiology,   & enantiospecific uptake of nutrients \\
production                 & MHD             & \\
\botrule
\label{Toverview}\end{tabular}}\end{table}

We begin with some historical remarks highlighting the significance of
helicity of magnetic fields (\Sec{HistoricalRemarks}).
In \Secss{ExternalFactors}{Spontaneous}, we discuss aspects of the three
types in more detail and then conclude in \Sec{Conclusions} with some
additional reflections.
We emphasize that we use the terms chirality, helicity, and handedness
rather interchangeably, although technically this is not always accurate.

\section{Historical remarks}
\label{HistoricalRemarks}

\subsection{Helicities in fluid dynamics and MHD}

In the context of fluid dynamics, the term helicity was coined by H.\ Keith
Moffatt,\cite{Mof69} who identified the topological equivalence between the
knottedness of vortex lines in fluid dynamics and the kinetic helicity.
In fact, the term helicity was already used by Robert Betchov\cite{Betchov61}
in 1961, but Moffatt proposed this name in his 1969 paper on the grounds
that this term is also used in subatomic physics to describe the alignment
or anti-alignment of spin and momentum of fermions, for example.
Mathematically, the mean kinetic helicity density is defined as
$\bra{\oo\cdot\uu}$, where the vorticity $\oo=\nab\times\uu$ is
the curl of the velocity $\uu$.
Kinetic helicity is a pseudoscalar, i.e., it changes its sign when the
system is inspected through a mirror.
Likewise, $\oo$ is a pseudovector, so it is more meaningful to plot it with
its sense of rotation (which changes in a mirror), rather than a vector with
an arrow at its end.

In the magnetic context, the corresponding quantity that we now call
magnetic helicity was already studied by Lodewijk Woltjer\cite{Wol58}
in 1958 to characterize force-free magnetic fields and by John Bryan
(J.B.) Taylor\cite{JBTaylor74} in 1974 to describe the relaxation of a
toroidal plasma.

\begin{figure}[t]
\begin{center}
\includegraphics[width=\textwidth]{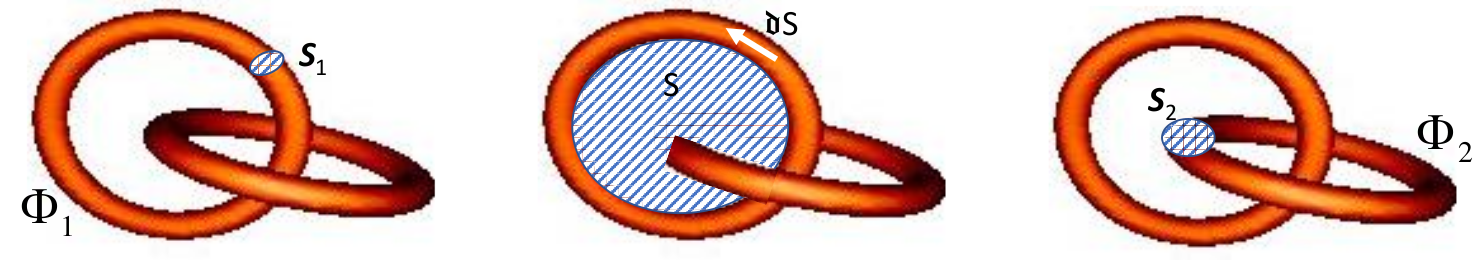}
\end{center}
\caption{Illustration showing that the line integral
$\oint_{\partial S}\AAA\cdot\dd\eell$ along flux tube $\Phi_1$,
can be written as a surface integral over the enclosed surface $S$,
$\int_S(\nab\times\AAA)\cdot\dd\SSS$, where $\partial S$
is the line along flux tube $\Phi_1$.
However, the only nonvanishing contribution comes from
flux tube $\Phi_2$, i.e., $\int_{S_2}\BB\cdot\dd\SSS$.}
\label{interlocked2b}
\end{figure}

\subsection{Magnetic helicity conservation}

The quantities that we shall focus on here are mainly the mean magnetic
helicity density $\bra{\AAA\cdot\BB}$, where $\BB=\nab\times\AAA$ is the
magnetic field expressed in terms of its vector potential $\AAA$, and
the magnetic helicity over a volume containing two interlocked flux loops,
$\int_{V_1+V_2}\AAA\cdot\BB\,\dd V=\int_{V_1}\AAA\cdot\BB\,\dd V
+\int_{V_2}\AAA\cdot\BB\,\dd V$.
The volumes $V_1$ and $V_2$ are those of the two tubes.
For $V_1$, we can split the volume integral into a line integral
$\oint_{\partial S}\AAA\cdot\dd\eell$ along the flux tube and
a surface integral $\int_{S_1}\BB\cdot\dd\SSS$ across the tube, i.e.,
\EQ
\int_{V_1}\AAA\cdot\BB\,\dd V
=\left(\int_{S_1}\BB\cdot\dd\SSS\right)
\left(\oint_{\partial S}\AAA\cdot\dd\eell\right);
\EN
see \Fig{interlocked2b}.
Using Stokes' theorem, the line integral along tube~1 can be rewritten
as a surface integral over the magnetic field going through the
flux ring, but its only nonvanishing contribution comes from the
other intersecting tube with surface $S_2$.
Thus, we have $\int_{V_1}\AAA\cdot\BB\,\dd V=\pm\Phi_1\Phi_2$,
and likewise for $V_2$, so we get a factor 2.
The sign depends on the relative orientation of
the field vectors in the two tubes, so we write
here $\int\AAA\cdot\BB\,\dd V=\pm2\Phi_1\Phi_2$.

The magnetic field is a pseudovector.
Its evolution is governed by the homogeneous Maxwell equation
$\partial\BB/\partial t=-\nabla\times\EE$, where $t$ is time and $\EE$
is the electric field.
Since $\nab\cdot\BB=0$, it is convenient to consider the uncurled
induction equation
\begin{equation}
\partial\AAA/\partial t=-\EE-\nab\phi,
\end{equation}
where $\phi$ is the electric (or scalar) potential.
To obtain an evolution equation for $\AAA\cdot\BB$, we compute
$\AAA\cdot\dot{\BB}+\dot{\bm{A}}\cdot\BB$, where dots denote partial time
differentiation, and find
\begin{equation}
\frac{\partial}{\partial t}\left(\AAA\cdot\BB\right)=-2\EE\cdot\BB
-\nab\cdot(\underbrace{\phi\BB+\EE\times\AAA}_{\mbox{helicity flux}}).
\end{equation}

We see that the magnetic helicity production depends on
$\EE\cdot\BB$ and on the presence of magnetic helicity fluxes,
$\FFF_H=\phi\BB+\EE\times\AAA$.
The $\EE\cdot\BB$ term plays important roles in electrically
non-conducting environments, for example during inflation in the early
universe when space-time became extremely diluted.
By contrast, in the contemporary universe, and even in the space between
galaxy clusters, there is still sufficient conductivity so that the laws
of MHD apply, and not the vacuum equations for electromagnetic waves,
which apply during inflation.
In MHD, the electric field is given by $\EE=-\uu\times\BB+\JJ/\sigma$,
where $\JJ$ is the current density and $\sigma$ the electric conductivity.
We see that the first term does not contribute to $\EE\cdot\BB$.
Thus, the only contribution comes from $\JJ\cdot\BB/\sigma$, so we have
\begin{equation}
\frac{\partial}{\partial t}\bra{\AAA\cdot\BB}=-2\eta\bra{\JJ\cdot\BB},
\end{equation}
where $\eta=(\mu_0\sigma)^{-1}$ is the magnetic diffusivity, and
angle brackets denote volume averaging.
When the conductivity is large, $\eta$ is small, and
$\eta\bra{\JJ\cdot\BB}$ converges to zero like $\eta^{1/2}$ as $\eta\to0$.

The smallness of $\eta$ in many astrophysical settings makes
$\bra{\AAA\cdot\BB}$ nearly perfectly conserved,\cite{Wol58}
except for the presence of magnetic helicity fluxes.
Those vanish in homogeneous systems (e.g., in homogeneous helical
turbulence), but astrophysical dynamos are usually not homogeneous and
magnetic helicity fluxes, for example out of the star or between its
northern and southern hemispheres, are believed to play important roles
in astrophysical dynamos to alleviate some serious constraints\cite{BS05}
arising from the magnetic helicity conservation otherwise.

\subsection{Helicity-driven large-scale dynamos}

To understand the aforementioned constraint, we need to emphasize that
magnetic helicity is usually connected with the presence of kinetic
helicity, $\bra{\oo\cdot\uu}$.
It can lead to an electromotive force along the mean magnetic
field, $\meanBB$, of the form
\EQ
\overline{\uu\times\bb}=\alpha\meanBB-\etat\mu_0\meanJJ,
\label{AlphaEffect}
\EN
where $\uu=\UU-\meanUU$ and $\bb=\BB-\meanBB$ are fluctuations of
velocity and magnetic fields, and $\meanJJ=\nab\times\meanBB/\mu_0$
is the mean current density.
Here, overbars denote averaging (for example planar or $xy$ averaging),
but the type of averaging depends on the particular problem.
Under the assumption of isotropy and high conductivity,\cite{KR80} we
have $\alpha=-\bra{\oo\cdot\uu}\tau/3$ and $\etat=\bra{\uu^2}\tau/3$,
and $\tau$ is the correlation time.
The $\alpha\meanBB$ term can lead to an exponential growth of $\meanBB$.
The second term just leads to an enhancement of the microphysical
magnetic diffusion, $\eta\mu_0\meanJJ$.

The $\alpha$ effect leads to the generation of magnetic helicity of
the mean field with $\meanAA\cdot\meanBB=O(\meanBB^2/k_{\rm m})\neq0$,
where $k_{\rm m}$ is the typical wavenumber of the magnetic field.
However, since the total magnetic helicity is conserved (and vanishing
of the field was very small initially), we must generate small-scale
magnetic helicity, $\bra{\aaaa\cdot\bb}$, of opposite sign.
The resulting current helicity,
$\bra{\jj\cdot\bb}\approx\kf^2\bra{\aaaa\cdot\bb}$, with some typical
wavenumber $\kf$ characterizing the fluctuation scales, quenches the
$\alpha$ effect and leads to slow dynamo saturation.\cite{Bra01}

We will not go into further details here, but refer the reader to
reviews on the subject.\cite{BS05,Rin19}
This field of research remains very active and there are still important
questions regarding magnetic helicity fluxes that remain controversial.

\section{Chirality driven by external factors}
\label{ExternalFactors}

\subsection{Examples of helicities: even and odd in $\BB$}

We discuss here three different helicities: kinetic helicity
$\bra{\oo\cdot\uu}$, magnetic helicity $\bra{\AAA\cdot\BB}$,
and cross helicity $\bra{\uu\cdot\BB}$.
The latter reflects the linkage between an $\oo$ tube (vortex tube)
and a $\BB$ tube (magnetic flux tube).
To understand the production of kinetic helicity, one has to realize
that it is a pseudoscalar.
This means, that it is the product of a polar vector and an axial vector.
Rotation, for example, is an axial vector, but that alone cannot
produce magnetic or kinetic helicity.
However, in the presence of both rotation $\OO$ and gravitational
stratification characterized by the gravitational acceleration $\grav$,
which is a polar vector, we can produce the pseudoscalar $\grav\cdot\OO$.
Thus, kinetic helicity can be produced if one can identify external factors
such as the combined presence of $\grav$ and $\OO$ that could
explain the presence of a non-vanishing helicity.
However, these external factors must also be even in the magnetic
field, so a nonvanishing $\bra{\uu\cdot\BB}$ with a systematic sign
cannot be explained in that way.
Finite cross helicity can, however, be explained if rotation was
replaced by an imposed magnetic field $\BB_0$, so that $\grav\cdot\BB_0$
would be finite.

We thus see that $\bra{\oo\cdot\uu}$ may be linked to $\grav\cdot\OO$,
although this needs to be (and has been) verified using a detailed
calculation.\cite{KR80}
In spherical coordinates $(r,\theta,\phi)$, where $\theta$ is colatitude
and the unit vectors of $\grav$ and $\OO$ are $\hat{\grav}=(-1,0,0)$
and $\hat{\OO}=(\cos\theta,-\sin\theta,0)$, respectively, we have
$\grav\cdot\OO=-\cos\theta$, which is negative in the north and positive
in the south.
This is indeed consistent with the observed sign of $\bra{\oo\cdot\uu}$,
and it is also found to govern the sign of the magnetic helicity, but
only at small and moderate length scales, i.e., $\bra{\aaaa\cdot\bb}$.

Regarding the cross helicity, there is indeed a systematic large-scale
magnetic field $\BB_0$ at the solar surface, although there can be
several sign changes in each hemisphere.
(The radius of the Sun is $700\Mm$, and $1\Mm=1000\km$, a useful unit
in solar physics!)
On smaller scales of $\ge20\Mm$, it also changes between the two sides
of a sunspot pair, which is consistent with observations.\cite{ZB18}
It turns out that $\bra{\uu\cdot\BB}\approx-(\etat/\cs^2)\,\grav\cdot\BB_0$,
where $\etat$ is the turbulent magnetic diffusivity and $\cs$ is the
sound speed.\cite{RKB11}
This production mechanism of cross helicity may play a role in theories
of shallow sunspot formation.\cite{BKR13,BGJKR14}

\subsection{Observing helicity}

\subsubsection{Measuring magnetic helicity from solar magnetograms}

Above the surface of the Sun, one often sees twisted structures
in extreme ultraviolet and x-ray images, which are suggestive of
helicity.\cite{Demoulin+Priest89,Gibson+02,Guo+13}
Even space observations of the surface of the Earth reveal cyclonic
cloud patterns that have opposite orientation in the northern and
southern hemispheres.
However, to draw a connection with magnetic helicity, one must first
detect the magnetic field.
This is possible through the Zeeman effect, which causes circular
polarization proportional to the line-of-sight magnetic field, and linear
polarization related to the perpendicular magnetic field component --
except for a $\pi$ ambiguity, which means that polarization measurements
are never able to tell where the tip of the magnetic field vector is,
so there is an uncertainty with respect to $180\degr$.
For strong enough magnetic fields, a ``disambiguation procedure''
based on a minimal magnetic energy assumption allows one to determine
the full $\BB$ vector at the solar surface.\cite{Georgoulis+05,
Rudenko+Anfinogentov14}
However, there is still not enough information about the changes of
polarization parameters along the line of sight, i.e., below and above
the surface of the Sun.

To make progress, one has to make some extra assumptions.
One possibility is to determine just the components normal to the surface,
$B_\|\equiv B_z$ and $J_\|\equiv J_z=\partial_x B_y-\partial_y B_x$,
where local Cartesian coordinates $(x,y,z)$ have been employed, and
compute $J_\|B_\|$.
This was first done by Seehafer,\cite{See90} who found that $J_\|B_\|$
is negative in most of the active regions in the northern hemisphere of
the Sun, and positive in most of the active regions in the southern one.
From $J_\|$ and $B_\|$, one can also determine a proxy of the magnetic
helicity spectra, $\tilde{A}_\|(\kk_\perp)\tilde{B}^*_\|(\kk_\perp)$,
where tildes denote Fourier transformation, $\kk_\perp$ is the wavevector
in the horizontal ($xy$ surface) plane, and the asterisk denotes the
complex conjugate.
Here, $\tilde{A}_\|=\tilde{J}_\|/\kk_\perp^2$.
This has been done\cite{ZBS14,ZBS16} and magnetic energy and helicity
spectra are shown in \Fig{phelicity_plat4b_Lidingo}(a) for active region
AR~11158.
Subsequent work\citep{ZB18} also presented cross helicity spectra.

\subsubsection{Magnetic helicity spectra from a time series}

A completely different approach, due to Matthaeus and et
al.,\cite{Matthaeus82} which also makes use of Fourier transformation,
is to use {\em in situ} observations of time series of the $\BB$ vector
in the solar wind at one point in space.
One can then make use of what is known as the Taylor hypothesis
to associate the temporal changes with different positions through
$\rr=\rr_0-\vv t$, where $\vv$ is the velocity vector of the solar wind,
and $\rr_0$ is a reference position.
Assuming homogeneity, i.e., that the statistical properties are
independent of position, one can write the magnetic two-point correlation
tensor in Fourier space as
\EQ
4\pi\bra{\tilde{B}_i(\kk)\tilde{B}_j(\kk)}=
(\delta_{ij}-\hat{k}_i\hat{k}_j)2\mu_0\EM(k)
-\ii\epsilon_{ijk}\hat{k}_k\HM(k),
\EN
where $\EM(k)$ is the magnetic energy spectrum, normalized such that
$\int\EM(k)\,\dd k=\bra{\BB^2}/2\mu_0$, and $\HM(k)$ is the magnetic
helicity spectrum with $\int\HM(k)\,\dd k=\bra{\AAA\cdot\BB}$.
This procedure revealed a clear hemispheric antisymmetry with respect
to the solar equator.\cite{BSBG11}

\begin{figure}[t]
\begin{center}
\includegraphics[width=\textwidth]{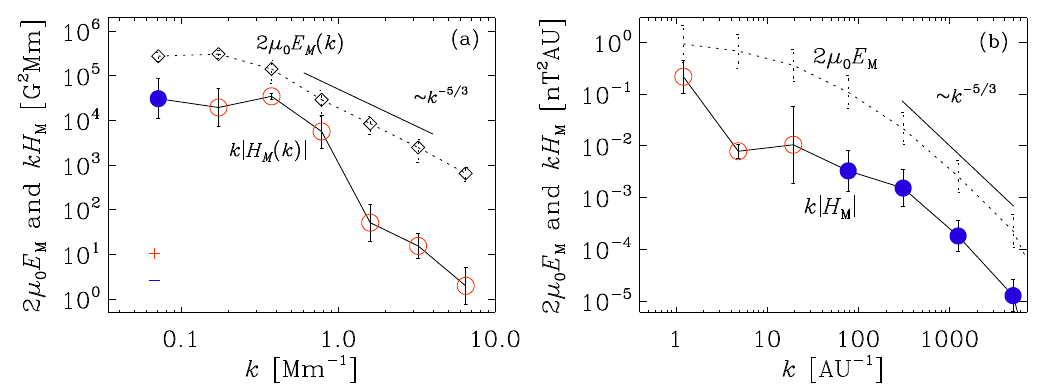}
\end{center}
\caption{
Magnetic energy and magnetic helicity spectra for southern latitudes
(a) at the solar surface in active region AR~11158, and
(b) in the solar wind at $\sim1\AU$ distance ($1\AU\approx149,600\Mm$).
Positive (negative) signs are shown as red open (blue filled) symbols.
Positive signs are the solar surface at intermediate and large $k$
correspond to positive values in the solar wind at small $k$.
Note that $1\G=10^{-4}\T=10^5\nT$.
}\label{phelicity_plat4b_Lidingo}
\end{figure}

Comparing \Fig{phelicity_plat4b_Lidingo}(a) and (b), which are here
both for the southern hemisphere, we see that at the solar surface,
$\HM(k)$ has the expected sign at small and intermediate scales
($k>0.1\Mm^{-1}$).
In the solar wind,\footnote{The mean distance between the Sun and the
Earth is one astronomical unit ($1\AU\approx149,600\Mm$).} however,
a positive sign is only seen at very large scales
($k<30\AU^{-1}\approx0.0002\Mm^{-1}$).
The typical wavenumber at which the spectral helicity changes sign at
the solar surface (radius $700\Mm$) is $k\approx0.1\Mm^{-1}$.
Expanding this linearly to a distance of $1\AU$ yields a corresponding
wavenumber of $k=70\AU^{-1}$, which is indeed where the spectrum in the
solar wind changes sign; see \Fig{phelicity_plat4b_Lidingo}(b).
However, the change is here the other way around: from positive to
negative at large $k$.
The reason for this sign mismatch is not yet fully understood, but it
is worth noting that a sign reversal has also been seen in idealized
numerical simulations of stellar and galactic dynamos embedded in
a turbulent exterior.\cite{BCC09, WBM11a, WBM11b}
Chiral solar wind MHD turbulence has also been studied in the equatorial
plane, but then both signs of helicity are possible.\cite{Zhu+14,Zhu17}
Both signs of magnetic helicity have also been found from multi-spacecraft
measurements in close proximity of the equatorial plane.\cite{Narita09}

\subsubsection{Canceling Faraday depolarization with helicity}

There is an intriguing possibility to determine magnetic helicity
from the cancelation of Faraday depolarization.\cite{BS14,HF14}
It might be particularly suitable for calculating the magnetic
helicity in the outskirts of edge-on galaxies, where one might
see the sign of magnetic helicity being reversed.

Normally, in the absence of magnetic helicity, Faraday rotation
causes Faraday depolarization.
This is caused by the superposition of different polarization
planes from different depths along the line of sight.
At the same time, however, the perpendicular component of the magnetic
field itself can rotate about the line of sign if it is helical.
These two effects can then either enhance each other (and make
the Faraday depolarization more complete), or they can offset
each other and lead to increased transmission or reduced
depolarization.

Mathematically, this can be seen by considering the observable
polarization, written in complex form as
\EQ
P(x,z,\lambda^2)\equiv Q+\ii U=p_0\int_{-\infty}^\infty \epsilon\,
e^{2\ii(\psi_P+{\phi}\lambda^2)}\,\dd y,
\label{Pint}
\EN
where $\psi_P=\psi_B+\pi/2$ is the electric field angle,
$\psi_B={\rm atan}(B_y,B_x)$ is the magnetic field angle,
$\epsilon(x,y,z)$ is the emissivity,
$p_0$ is the degree of polarization,
$\lambda$ is the wavelength,
\EQ
\phi(x,y,z)=-K\int_{-\infty}^y n_{\rm e}(x,y',z)\,B_\|(x,y',z)\,\dd y'
\label{phiz}
\EN
is the Faraday depth, with $n_{\rm e}$ being the electron density
and $K=0.81\m^{-2}\cm^3\uG^{-1}\pc^{-1}=2.6\times10^{-17}\G^{-1}$
being a constant.
Evidently, Faraday depolarization is canceled if
$\psi_P+{\phi}\lambda^2=0$.

In essence, for edge-on galaxies, we expect maximum polarized emission
in diagonally opposite quadrants of a galaxy; see Fig.~19.12 of
Ref.~\citenum{Bra15}.
This technique has also been applied to synthetic data of the solar
corona.\cite{BAJ17}

\subsubsection{Magnetic helicity proxy}

In the context of cosmology, a proxy of parity breaking and finite
helicity of the cosmic microwave background and the Galactic foreground
is obtained by decomposing the observed linear polarization in the sky
into parity-even and parity-odd contributions.
This is done by expanding the complex polarization $P\equiv Q+\ii U$
with Stokes parameters $Q$ and $U$ into a spin-2 spherical harmonics
as\citep{Kamion97,SZ97}
\EQ
\tilde{R}_{\ell m}=\int_{4\pi}
(Q+\ii U)\,_2 Y_{\ell m}^\ast(\theta,\phi)\,
\sin\theta\,\dd\theta\,\dd\phi.
\label{EBfromQU}
\EN
The parity-even ($E$) and parity-odd ($B$) contributions are obtained as
the real and imaginary parts of the return transformation as\cite{Dur08,
KK16}
\EQ
E+\ii B\equiv R=\sum_{\ell=2}^{N_\ell}\sum_{m=-\ell}^{\ell}
\tilde{R}_{\ell m} Y_{\ell m}(\theta,\phi).
\EN

\begin{figure}[t]
\begin{center}
\includegraphics[width=\textwidth]{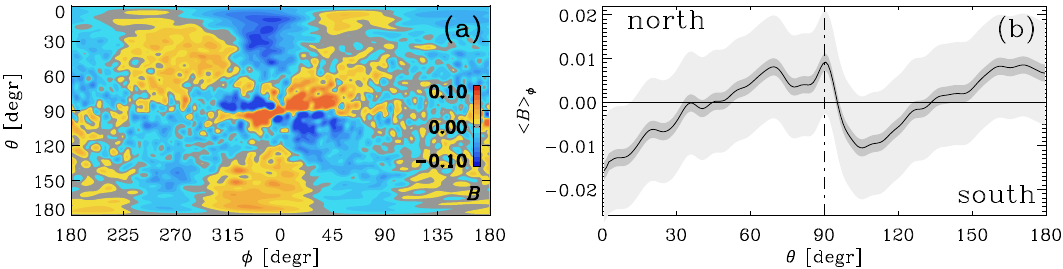}
\end{center}
\caption{Left: Galactic $B$ mode polarization.
Right: longitudinally averaged $B$ mode polarization.
Here, $\theta$ and $\phi$ are Galactic colatitude
($=90\degr-\mbox{latitude}$) and longitude.
}\label{RfromP_shift_last}
\end{figure}

In cosmology, one usually considers correlations between $E$ and $B$,
as well as temperature and $B$, for example.
However, it is also useful to consider $B$ on its own, as was done
in the context of the solar magnetic field\cite{Bra19} and in
the context of the Galactic magnetic field.\cite{BB20b}
The result for our Galaxy is shown in \Fig{RfromP_shift_last}.

We see that the $B$ polarization is locally antisymmetric about the
equatorial plane, but there are canceling contributions from different
longitudes, so the longitudinal average is much smaller.
Nevertheless, even the longitudinal average does show a hemispheric
antisymmetry.

Comparing with synthetic observations using the magnetic field from
idealized dynamo models\cite{BF20} shows that this hemispheric dependence
does not originate from the different signs of the magnetic helicity
expected in the northern and southern hemispheres, but from the spiral
pattern of our Galaxy, where the views from the north and south correspond
to mirror images of each other.

\section{Magnetic helicity throughout the whole Universe}
\label{WholeUniverse}

There is at present no definitive observation of finite helicity
throughout all of the Universe, but the possibility certainly
exists.\citep{V01}
In this section, we first discuss two quite different mechanisms.
One is related to the chiral magnetic effect (CME) and the other to
inflationary magnetogenesis.
Both generate helical magnetic fields, and the electromagnetic stress can
generate relic gravitational waves (GWs) that are circularly polarized.
Such waves, once generated, would not dissipate and would only dilute
under the cosmic expansion.
They could still be observable with space interferometers\cite{Caprini+16,
2017arXiv170200786A, Taiji} and with pulsar timing
arrays.\cite{NANOGrav2020}
Measuring circular polarization could provide a clean mechanism
for determining the sign of helicity in the Universe.

\begin{figure}[t]
\begin{center}
\includegraphics[width=\textwidth]{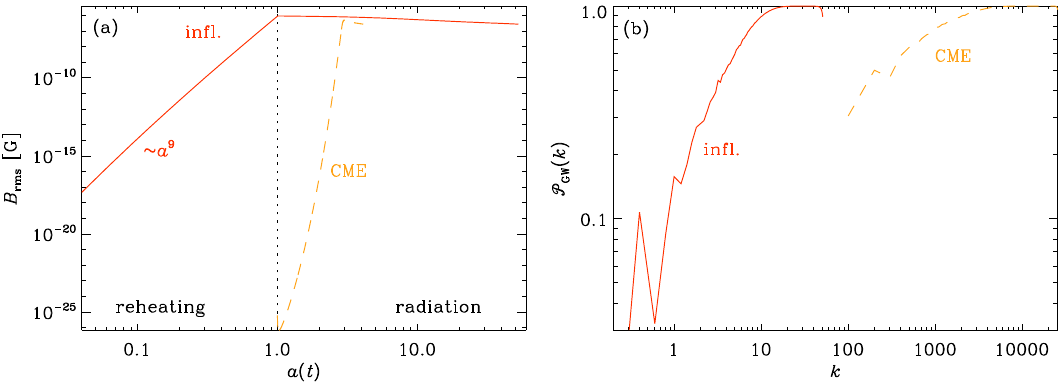}
\end{center}
\caption{(a) Growth of magnetic field from the CME (orange) and
inflationary magnetogenesis (red), where time is here expressed in terms
of the scale factor of the Universe $a(t)$, which increases monotonically
and is set to unity at the beginning of the radiation-dominated era.
Note the algebraic increase $\Brms\propto a^9$.
(b) Circular polarization spectra of GWs produced from the CME (orange)
and inflationary magnetogenesis (red).
}\label{pcomp_befaftP_ppol_comp_Lidingo}
\end{figure}

In the following, we describe the two generation mechanisms
and compare in \Fig{pcomp_befaftP_ppol_comp_Lidingo} the growth
of the resulting magnetic field and the circular polarization
of GWs that could be observed in future using
space interferometers in the millihertz range.\cite{Caprini+16,
2017arXiv170200786A, Taiji}

\subsection{The CME}

The CME is a quantum effect associated with the systematic alignment of
the spin $\sss$ of electrically charged fermions (electrons or positrons,
for example) with the momentum $\pp$.
A nonvanishing net helicity $\sss\cdot\pp$ originates from the
parity-breaking weak force and manifests itself in the $\beta$ decay,
for example, where spin and momentum of electrons are antialigned, i.e.,
$\sss\cdot\pp<0$.
The sign would be opposite for antimatter, i.e., for positrons, for
example.
However, at low energies, spin flipping occurs,\cite{Boyarsky+21}
making this effect important only for highly relativistic plasmas at
high enough temperatures.

In the presence of a magnetic field, the spin of chiral fermions aligns
itself with the magnetic field, and, owing to the finite momentum and
charge of the fermions, they produce a net current along the magnetic
field,\cite{Vilenkin80, JS97, BFR12, BFR15, Roga_etal17}
\EQ
\JJ=24\alpha_{\rm em}\,(n_{\rm L}-n_{\rm R})
\left(\frac{\kB T}{\hbar c}\right)^2\BB,
\EN
where $\alpha_{\rm em}\approx1/137$ is the fine structure constant,
$n_{\rm L}$ and $n_{\rm R}$ are the number densities of left- and
right-handed fermions, respectively, $\kB$ is the Boltzmann constant,
$T$ is the temperature of the Universe today, $\hbar$ is the reduced
Planck constant, and $c$ is the speed of light.

We see the analogy with \Eq{AlphaEffect}, where we had a mean
electromotive force $\meanEMF\equiv\overline{\uu\times\bb}$ with a mean
current $\meanJJ=(\alpha/\etat\mu_0)\,\meanBB$ produced along the mean
magnetic field for turbulent for helical turbulence.
Similar to the dynamo effect in helical turbulence, there is a dynamo
effect associated with chiral plasmas.
This was discussed in the context of cosmology\cite{JS97} as a mechanism
for producing magnetic fields in the early Universe.
This effect produces helical magnetic fields, and the associated
helicity reduces the net chirality of the fermions such
that the total chirality is conserved, i.e.,
\begin{equation}
  \label{eq:1}
  (n_{\rm L} - n_{\rm R}) + \frac{4\alpha_{\rm em}}{\hbar c}
  \langle \AAA\cdot \BB\rangle = \const.
\end{equation}

The length scales are small compared with the Hubble radius at any given
time and the resulting field strengths are limited to\cite{Bran_etal17}
\begin{equation}
  |\bra{\AAA\cdot\BB}|\la\xiM\bra{\BB^2}\la(0.5\times10^{-18}\G)^2\Mpc,
\end{equation}
where $\xiM$ is the magnetic correlation length, measured here in
megaparsec ($1\Mpc\approx3\times10^{24}\cm$).
This value of $\xiM\bra{\BB^2}$ is below the lower limits derived from
the non-observation of secondary GeV photons that are expected from
inverse Compton scattering of TeV photons from energetic blazers.
One can therefore not be very hopeful that those lower limits on the
magnetic field can be explained as a result of the CME.
However, stronger fields could still be produced during inflation,
as will be discussed next.

\subsection{Inflationary magnetogenesis}

Models of inflationary magnetogenesis tend to invoke conformal
invariance-breaking to generate magnetic fields as a result of stretching
without diluting the magnetic energy during the inflationary expansion
and the subsequent reheating phase when most of the relevant particles
and photons were produced.
Conformal invariance-breaking means that the term $F_{\mu\nu}F^{\mu\nu}$
in the Lagrangian density is replaced by $f^2F_{\mu\nu}F^{\mu\nu}$
with $f\neq1$ during inflation\cite{Ratra92} and reheating.\cite{DR12}
In the presence of such a term, the vacuum evolution equation
for the vector potential changes from a standard wave equation
$(\partial^2/\partial t^2+k^2)\tildeAA=0$ to $(\partial^2/\partial
t^2+k^2-f''/f)(f\tildeAA)=0$, where $t$ is now conformal time and the
primes on $f$ denote derivatives with respect to conformal time.

Commonly adopted forms of $f$ include $f\propto a^\alpha$ during inflation
and $f\propto a^{-\beta}$ during reheating.\cite{Ferreira+13, Sharma+17,
Sharma+18, Bran+Shar21}.
In the presence of pseudoscalars $\gamma$, such as axions, one also
expects terms of the form $\gamma f^2F_{\mu\nu}\tilde{F}^{\mu\nu}$
in the Lagrangian density.
The evolution equation for $\tildeAA$ takes then the form
\begin{equation}
\left(\frac{\partial^2}{\partial t^2}+k^2\pm2\gamma k\frac{f'}{f}
-\frac{f''}{f}\right)(f\tildeA_\pm)=0,
\end{equation}
where $\tildeAA=\tildeA_+\eee_++\tildeA_-\eee_-$ has been expressed
in terms of the polarization basis 
$\tildeee_\pm(\kk)=[\tildeee_1(\kk)\pm\ii\tildeee_2(\kk)]/\sqrt{2}\,\ii$
with $\ii\kk\times\tildeee_\pm=\pm k\tildeee_\pm$, and $\tildeee_1(\kk)$,
$\tildeee_2(\kk)$ represent units vectors orthogonal to $\kk$
and orthogonal to each other.
\FFig{pcomp_befaftP_ppol_comp_Lidingo}(a) shows the resulting algebraic
growth of the magnetic field in comparison with the exponential growth
from the CME, and panel (b) shows the circular polarization spectrum
${\cal P}_{\rm GW}(k)$.
It is defined as\cite{RoperPol+20b}
\EQ
\!\!{\cal P}_{\rm GW}(k) = \!\left.\int\!2\,\Imag\,\tildeh_+\tildeh_\times^*\,k^2\dd\Omega_k
\right/\!\!\int\!\left(|\tildeh_+|^2+\tildeh_\times|^2\right) k^2\dd\Omega_k.
\EN
We see that the degree of polarization reaches nearly 100\% in a certain
range. \cite{BHKRS21, Bran+He+Shar21}

\subsection{Detecting handedness from unit vectors in the sky}

In addition to measuring the polarization of GWs as an indicator of
the helicity of the underlying magnetic field, there is yet another
interesting method that we describe here briefly.
Suppose we observed energetic photons from a particular astrophysical
source from three slightly different directions, $\nnn_1$, $\nnn_2$, and $\nnn_3$, 
and that their energies $E_i$ are ordered such that $E_1<E_2<E_3$,
then we can construct a pseudoscalar
\EQ
Q=(\nnn_1\times\nnn_2)\cdot\nnn_3.
\EN
The quantity $Q$ would change sign in a mirror image of our Universe.
The value of $Q$ may well be zero within error bars, but if it is not,
its sign must mean something.

Detailed calculations have shown the TeV photons from blazars (i.e.,
the accretion disk around supermassive black holes) can upscatter on
the cosmic background light and produce GeV photons.\cite{Tashiro+13}
Using about 10,000 photons observed with Fermi Large Area Telescope
data over a period of about five years, Tashiro et al.\cite{Tashiro+14}
found $Q<0$ for all possible photon triples in certain energy ranges.
They interpreted this as evidence in favor of a baryogenesis scenario that
proceeds through changes in the Chern-Simons number, which implies the
generation of magnetic fields of negative helicity;\cite{vachaspati2001}
see also Ref.~\citenum{tanmay2021} for a review.

In all the studies of $Q$ done since then,\cite{Chen+15, Chen+15b,
Long+15, Long+15b} one introduced a cutoff toward low Galactic latitudes
so as to avoid excess contamination from our Galaxy.
Using synthetic data, it has been found\cite{AJB20} that it is this
procedure that leads to the occurrence of large statistical errors in
the estimate of $Q$.
Updated observations covering 11 years turned out to be no longer
compatible with a detection of a negative value of $Q$.\cite{AJB20} 
This has been confirmed in subsequent work.\cite{Kachelriess+20}
Thus, although this method could work in principle, it would require
much better statistics.

In this connection, it is interesting to note that an equivalent quantity
$Q$ can be determined for a variety of different observations.
Suppose there are three sunspots of different strengths on the surface
of the Sun, this again implies a finite handedness.
We can then ask whether this handedness could be linked to the helicity
of the underlying magnetic field.
Idealized model calculations have shown that this is indeed the
case.\cite{BB18}

\section{Spontaneous chirality production}
\label{Spontaneous}

\subsection{Biological homochirality}

In astrobiology, an important question concerns the origin of biological
homochirality.\cite{Rothery:2011}
In solution, many organic molecules tend to rotate the polarization plane
of linearly polarized light.
One refers to the substances as either levorotatory ({\sc l} for
left-handed) or dextrorotatory ({\sc d} for right-handed).
Almost all amino acids of terrestrial life are of the {\sc l} form,
and almost all sugars are of the {\sc d} form, for example the sugars
in the phosphorus backbones of deoxyribose nucleic acid (DNA).
The origin of this homochirality can be explained in terms of two
essential processes: autocatalysis and mutual antagonism -- an old
idea that goes back to a paper by F.\ C.\ Frank\cite{Frank} of 1953.
Interestingly, this is the same year when Watson and
Crick\cite{Watson+Crick53} discovered the helix structure of DNA.

Autocatalysis produces ``more of itself'', i.e., it can catalyze the formation
of chiral molecules of the {\sc l} form from an achiral substrate $A$
in the presence of {\sc l} and, conversely, it can catalyze the formation
of molecules of the {\sc d} form in the presence of {\sc d}.
The corresponding reactions
\EQ
L+A \to 2L
\quad\mbox{and}\quad
D+A \to 2D,
\EN
with rate coefficient $k_{\rm C}$,
imply that the associated concentrations $[L]$ and $[D]$ obey
\EQ
\frac{\dd}{\dd t}[L]=k_{\rm C}[A][L]-...
\quad\mbox{and}\quad
\frac{\dd}{\dd t}[D]=k_{\rm C}[A][D]-...,
\EN
which leads to exponential growth with time $t$ of both
$[L]=[L]_0 e^{k_{\rm C}[A]t}$ and $[D]=[D]_0 e^{k_{\rm C}[A]t}$,
with initial values $[L]_0$ and $[D]_0$.
However, this process alone does not change the enantiomeric excess (e.e.),
$\mbox{e.e.}=([L]-[D])/([L]+[D])$, which will always be equal to the
initial value.
This is because we still need mutual antagonism, which
will be explained next.

For a long time, it remained unclear what would correspond
to Frank's mutual antagonism.
The relevant understanding was put forward by Sandars.\cite{San03}
At that time, it was thought that homochirality was a prerequisite
to the origin of life.
This idea was based on an experimental result by Joyce et
al.,\citep{Joyce84} which showed that in template-directed polymerization
of oligomers of one chirality, polymers of the opposite chirality
terminate further polymerization.
This was called enantiomeric cross-inhibition, and was regarded
as a serious problem for the origin of life, and that life could only
emerge in a fully homochiral environment.\cite{GK89}
Sandars realized that enantiomeric cross-inhibition could just be the
crucial mechanism that corresponds to Frank's mutual antagonism that
would lead to the emergence of homochirality.
The corresponding reaction and rate equations,
with rate coefficient $k_{\rm I}$, are
\EQ
L+D \to 2A
\quad\mbox{and}\quad
\frac{\dd}{\dd t}[A]=2k_{\rm I}[L][D]-...\,.
\EN

Multiple extensions of Sanders' model have been
produced\cite{Saito+Hyuga03, BAHN, Wattis+05, Gleiser+Walker08} and there
are also other variants that are not based on nucleotides, but
on peptides.\cite{Plasson+04, BLL07, Gleiser+Walker09, Konstantinov+20}
If one regards these first polymerization reactions as the first steps
toward life, one could then say that homochirality emerges as a consequence
of life, and not as a prerequisite.\cite{San05, NeubeckReview}

The lessons learnt from astrobiology may well be applicable to other
fields of physics, and in particular to MHD.
Examples were found in the context of the magnetobuoyancy
instability\cite{Chatter11} and the Tayler
instability.\cite{Gellert+11,Bonanno+12}
In those cases, there are two unstable eigenfunctions that are helical
and have positive and negative helicities, respectively, but their growth
rates are equal.
This process corresponds to autocatalysis.
The nonlinearity in the MHD equations, associated with the Lorentz force,
corresponds to mutual antagonism.
The resulting amplitude equations in MHD are the same as in the production
of homochirality.\cite{Bonanno+12}
Another more recent example of this type has been found in studies
of the CME when the chiral chemical potential is fluctuating around
zero.\cite{Schober21a,Schober21b}
Again, one chirality becomes eventually dominant, and this choice depends
on details in the initial conditions.

\subsection{Spatially distinct domains of chirality}

\begin{figure}[t]
\begin{center}
\includegraphics[width=\textwidth]{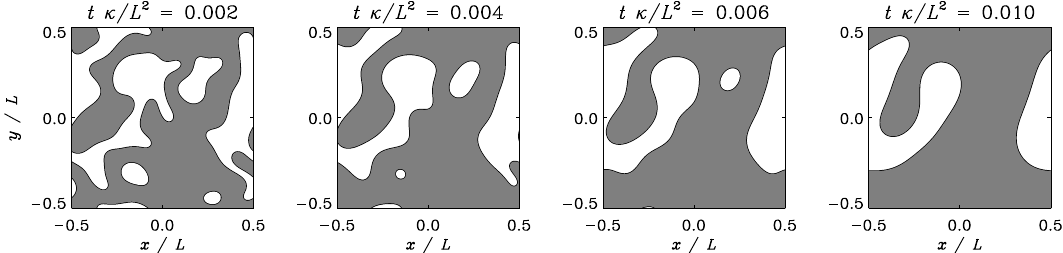}
\end{center}
\caption{Gradual shrinking of isolated islands on one handedness
(white), leading to the eventual dominance of the other (gray).
}\label{pxxx_chiral_Lidingo}
\end{figure}

In spatially extended domains, the evolution equations for $[L]$ and
$[D]$ attain additional spatial diffusion terms and become then similar
to the Fisher equation\cite{FisherEqn} which describes front propagation.
In the present case, once fronts of {\sc l} and {\sc d} polymers come
into contact, the front between them comes to a halt and cannot propagate
any further, unless the front is curved.
If it is curved, the front continues to propagate in the direction of largest
curvature.\cite{BM04}
This leads in the end to small near-circular islands that shrink until
they disappear; see \Fig{pxxx_chiral_Lidingo}.
This means that the enantiomeric excess changes in a piecewise linear
fashion.

\subsection{Analogous mechanisms in other systems}

Given that closed patches of one handedness always shrink and eventually
disappear, it is clear that the dominant chirality must in the end be
that of the outside of the last surviving patch.
Thus, it is not necessarily the one that was initially the most dominant
one.

\begin{figure}[t]
\begin{center}
\includegraphics[width=\textwidth]{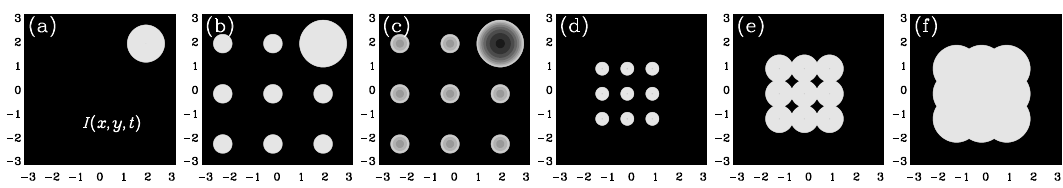}
\end{center}
\caption{The number of infected $I(x,y)$ in a two-dimensional
Cartesian plane $(x,y)$, as obtained from model
calculations,\cite{Brandenburg20piecewise} showing
(a) a circular spreading center in the upper right corner,
(b) the subsequent emergence of eight additional spreading centers that
(c) continue to grow, but with decreasing $I(x,y)$ in their centers
due to recovery or death.
Note that the total length of the periphery increases when the number
of spreading centers increases.
Panels (d)--(f) show a case where spreading centers merge, so the
growth in the length of the periphery declines.
}\label{psav_Lidingo}
\end{figure}

\begin{figure}[t]
\begin{center}
\includegraphics[width=\textwidth]{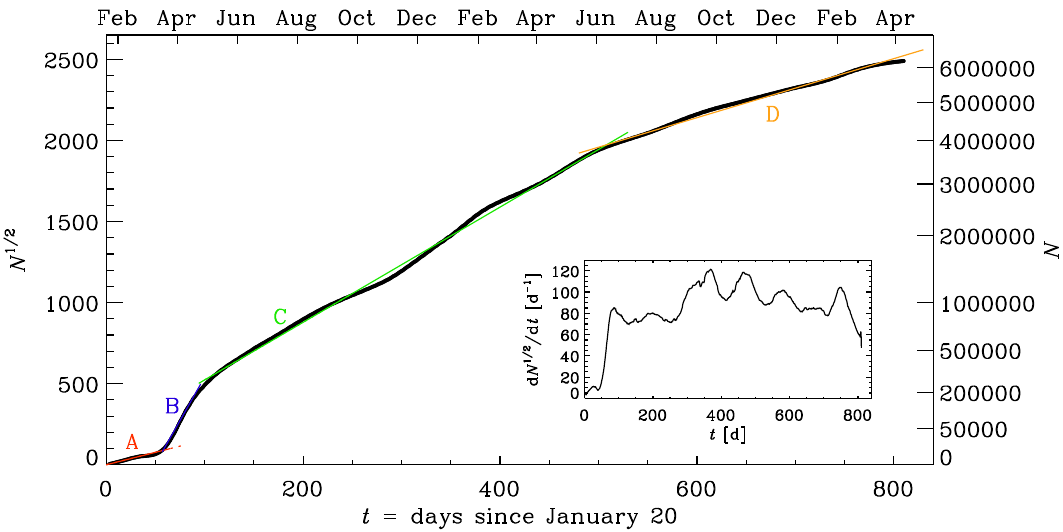}
\end{center}
\caption{Square root of the number $N$ of deaths, which is regarded as a
proxy of the number of infected that is more reliable than the reported
number of SARS-CoV-2.
Note the piecewise linear growth in $N^{1/2}$, corresponding to a
piecewise quadratic growth.
The line segments A--D are described in the text.
}\label{p3extn}
\end{figure}

A piecewise linear evolution is common to many spatially extended systems,
including those describing the spread of SARS-CoV-2 over the past two years.
Here, however, it is not the number $N$ itself, but its square root,
$N^{1/2}$.
In \Fig{psav_Lidingo}, we show the spatial geometry of a hypothetic
spreading center in the upper right corner.
The speed of growth depends just on the length of the periphery.
At a later time, there will be new spreading centers, so the total
length of the periphery increases.
This happened in the middle of March 2020; see \Fig{p3extn}.
During the first part of the epidemic (denoted by a red A), the evolution
was comparatively slow and the disease was essentially confined just
to China.
During the subsequent phase (denoted by a blue B) it spread all over
the world.
This led to an increase in the total length of the periphery.
Later, different spreading centers began to emerge, so the total
periphery has now decreased again, and the growth has slowed down
(denoted by the green and orange segments C and D, respectively),
but it always remained piecewise quadratic and was never
exponential.\cite{Brandenburg20piecewise}
The spreading of SARS-CoV-2 is obviously no longer directly related to
the topic of chirality in astrophysics, but it is interesting to see
that the mathematics of front propagation in spatially extended domains
is similar to that of left and right handed life forms invading the
early Earth.\cite{BM04}

\section{Conclusions}
\label{Conclusions}

In this work, we have sketched three rather different ways of achieving
chirality in astrophysics and astrobiology: externally driven,
intrinsically driven, and spontaneous chirality production.
The first two mechanisms are particularly relevant to fluid dynamics and
magnetic fields, while the last one of spontaneous chirality production is
mainly relevant to astrobiology and to the origin of life, but it remains
hypothetic until one is able to find an example of another genesis of
life independent of that on Earth, for example on Mars or on some of
the icy moons in our solar system.\cite{Rothery:2011}

The idea of propagating fronts of life forms of opposite handedness is
intriguing and it would be useful to reproduce this in the lab.
This may not be easy because such fronts propagate relatively rapidly
under laboratory conditions.
In fact, their propagation resembles the propagation of epidemiological
fronts, such as the black death\cite{Noble74} and perhaps even
SARS-CoV-2.\cite{Brandenburg20piecewise}
However, the application to the early Earth implies much larger spatial scales.
Earlier work quoted half a billion years as a relevant time scale.\cite{BM04}
However, if one thinks of the deep biosphere of Mars, it may not be
impossible to explain a possible detection of opposite chiralities of DNA,
if such should ever be observed on Mars or in its permafrost.

Alternatively, it is possible that the chirality of biomolecules is
determined through an external influence.\cite{GB20}
Such an influence would then, similarly to the intrinsically driven
chirality discussed in \Sec{WholeUniverse}, be related to the
parity-breaking weak force.
This possibility cannot easily be dismissed.
In particular, a 2\% enantiomeric excess in favor of the {\sc l} form has
been found for amino acids in the Murchison meteorite.\cite{Pizzarello+00}
This was a pristine meteorite rich in organics, as already evidenced by
the smell reported by initial eyewitnesses.
On the other hand, those molecules are also susceptible to contamination,
while those not susceptible to contamination did not show any enantiomeric
excess.

Determining a global chirality that is the same throughout the entire
Universe would be a major discovery.
Measuring the chirality through circular polarization of GWs would likely
be the most definitive proof of parity violation in the Universe.

\section*{Acknowledgments}

I thank Tanmay Vachaspati, J\'er\'emy Vachier, and Jian-Zhou Zhu for
comments and noticing some mistakes.
I am particularly grateful to Mats Larsson for having provided us with
an early opportunity for an in-person meeting in June 2021, right when
the quadratic growth of SARS-CoV-2 showed a break; see \Fig{p3extn}.
This work was supported in part through the Swedish Research Council,
grant 2019-04234.
I acknowledge the allocation of computing resources provided by the
Swedish National Allocations Committee at the Center for Parallel
Computers at the Royal Institute of Technology in Stockholm.

\bibliographystyle{ws-procs961x669}
\bibliography{ref}

\end{document}